\documentclass[conference]{IEEEtran}
\IEEEoverridecommandlockouts
\usepackage{cite}
\usepackage{amsmath,amssymb,amsfonts}
\usepackage{algorithmic}
\usepackage{graphicx}
\usepackage{textcomp}
\usepackage{xcolor}
\usepackage{subfigure}
\usepackage{geometry}
\def\BibTeX{{\rm B\kern-.05em{\sc i\kern-.025em b}\kern-.08em
    T\kern-.1667em\lower.7ex\hbox{E}\kern-.125emX}}
\begin{document}

\title{IoT Performance for Maritime Passenger Evacuation
\thanks{This research was partially supported by the Institute of Theoretical and Applied Informatics, Polish Academy of Sciences (IITIS-PAN), and partially by the National Natural Science Foundation of China (NSFC), Grant No. 51979216 and the Natural Science Foundation of Hubei Province, China, Grant Nos. 2021CFA001 and 20221j0059.}
}
	\author{\IEEEauthorblockN{1\textsuperscript{st}Yuting Ma}

	\IEEEauthorblockA{IITiS-PAN, Gliwice, PL,\\
		Wuhan University of Technology, China\\
		yma@iitis.pl}

\and \IEEEauthorblockN{2\textsuperscript{nd}Erol Gelenbe}

\IEEEauthorblockA{IITiS-PAN, Gliwice, PL,\\
Universit\'{e} C\^{o}te d'Azur, Nice, FR,\\
Ya\c{s}ar University, Bornova, TR\\
seg@iitis.pl}

\and
\IEEEauthorblockN{3\textsuperscript{rd}Kezhong Liu}

\IEEEauthorblockA{Wuhan University of Technology\\
Wuhan, China \\
kzliu@whut.edu.cn}
}

\maketitle

\begin{abstract}
The safe and swift evacuation of passengers from Maritime Vessels, requires effective Internet of Things (IoT) and information and communication technology (ICT) infrastructures. However, during emergencies, delays in IoT and ICT systems that provide guidance to evacuees can impair the effectiveness of the evacuation process. This paper presents simulations that explore the impact of this key aspect. The methodology builds upon the deadline-aware adaptive navigation strategy (ANT), offering at each decision step the path segment which minimizes the evacuation time for each evacuee. The simulations on a real cruise ship configuration, show that delays in the delivery of correct instructions to evacuees can significantly hinder the effectiveness of the evacuation service. These findings stress the need to design robust and computationally fast IoT and  ICT systems to support evacuation systems for ships, underscoring the key role played by IoT in the success of passenger evacuation. 
\end{abstract}


\section{Introduction}
In the context of emergencies, the effectiveness of evacuation methods is crucial for ensuring the safety of people and vehicles \cite{Wu2,chu2019emergency,luo2022they}. These methods rely on technologies encompassing sensing, communication, and signaling, which play pivotal roles in (a) locating individuals and vehicles, (b) communicating with them to relay information about ongoing conditions, and (c) guiding them along efficient pathways to safety.

Research in this field involves the utilization of sensing and communication technologies \cite{birajdar2020development}, crowd monitoring \cite{singh2021crowd}, hazard modeling and prediction \cite{choi2019optimal}, evacuation simulation \cite{Huibo}, and evacuation path planning \cite{ibrahim2016intelligent}. Specifically, offline simulation of evacuation strategies facilitates the design and comparison of sensing and communication technologies, as well as algorithms \cite{Huibo2} aiming to enhance or optimize the performance and robustness of evacuation strategies.

Emergency management simulation research addresses the movement of people in diverse scenarios, including sports arenas, tourist sites, leisure venues \cite{meschini2023bim,chu2019evacuation}, and large ships \cite{fang2023evacuation}, under unusual conditions or adversarial situations like facility breakdowns, fires, or panic.
With the global growth of the cruise ship industry, passenger safety in maritime trANTportation has gained increased attention \cite{liu2022survey}. Despite advanced accident prevention systems, maritime accidents still occur, underscoring the importance of effective evacuation methods for passenger ships \cite{sarvari2018studies}. Recent research in this area has seen considerable developments \cite{li2023data}.

Due to the difficulty of artificially creating or reproducing maritime emergencies, simulation of human evacuation on ships has become essential for designing both civilian and military naval vessels \cite{nasso2019simplified,wang2022numerical,arshad2022determinants}. Notably, the International Maritime Organization (IMO) has issued Circulars and guidelines for evacuation simulation in passenger ships \cite{international2016guidelines}. However, existing research and guidelines often overlook realistic features impacting human behavior during emergencies \cite{Huibo2,xie2022integrated}, such as variations in traversal times and the impact of technology-induced delays and message loss \cite{fang2022simulation,wang2023novel}. This paper investigates the influence of these unpredictable factors on the performance of contemporary simulation methods for vessel evacuation.

The paper employs a simulation framework to assess the impact of imperfections resulting from communication technologies, such as lost or delayed messages, and from passenger misunderstandings or panic during emergency management simulations. The tool utilized for this investigation is built on the AnLogic simulator \cite{niu2023emergency}, as detailed in Section \ref{SectionIII}.

The subsequent sections are organized as follows: Section \ref{SectionII} reviews related work on evacuation in passenger ships and built structures, Section \ref{SectionIII} provides a detailed description of the ship emergency evacuation simulation, and Section \ref{SectionIV} presents simulation results regarding the impact of communication system congestion. Finally, Section \ref{SectionVII} offers conclusions and suggestions for future research.

\subsection{Related work} \label{SectionII}

Emergency management systems guiding evacuees typically employ $\textbf{path-planning}$ algorithms, such as Dijkstra, A*, RRT, ANT Colony Optimization, Genetic algorithms, and others, to minimize evacuation delay or maximize the distance from hazard nodes to the exit node, utilizing 2D/3D maps of built environments \cite{Desmet, Kokuti2}. However, the resulting paths may become impractical as they include unexpected hazards like fire spread, flooding, or failures in emergency management systems. Efforts have been made to address dynamic dangers using the Expected Number of Oscillations (ENO) concept \cite{wang2014oscillation}, quantifying emergency dynamics and exploring paths with the smallest probability of frequent changes. Global path planning methods often require complete system specifications, contradicting realistic scenarios. Subsequent studies relax this requirement, allowing for path computation without precise prior knowledge of all hazard and exit information \cite{Filip, Huibo2}.

Methods such as Artificial Potential Fields and local neighborhood techniques \cite{tseng2006wireless, Kokuti}, possibly with partial reversal, help evacuees avoid hazardous areas. However, most studies overlook the necessity of providing directions ensuring that the "time needed to reach the exit" for each evacuee remains under a specified upper bound (e.g., ship survival or capsizing time) in worst-case scenarios. Thus, this paper adopts the ANT (Adaptive Navigation Strategy) evacuation algorithm \cite{ma2020ant}, explicitly incorporating a guaranteed exit deadline for each evacuee in each location. This deadline bound is derived from the International Maritime Organization (IMO) recommendations for ships.

\section{The Simulation framework} \label{SectionIII}

We present the simulation framework employed to assess the impact of computer and communication technology delays on providing routing instructions to evacuees within an event-driven simulation framework. This framework comprises two integral parts: (1)  AnyLogic Simulation Software \cite{niu2023emergency}: This software incorporates the layout of the physical environment where the evacuation takes place. AnyLogic utilizes a "pedestrian" software library employing a social force model, akin to social potential fields, for evacuee movement. (2) Path-Planning Module in Python: A path-planning module written in Python computes evacuation directions based on the ANT algorithm, an extension of the Rapid Routing with Guaranteed Delay Bounds algorithm \cite{baruah2018rapid} that was called ANT in prior work \cite{ma2020ant}. This module conveys computed instructions to the AnyLogic simulation software at each simulation step when movement instructions are updated.

The ANT algorithm is implemented in our simulator with the following features: 
\begin{itemize}
	\item  ANT moves evacuees along paths with minimal delay, mitigating harmful effects from dynamic hazards. The algorithm assumes knowledge of hazard propagation (velocity and direction), as well as typical and worst-case delay across each edge of each path.
\item As hazards progress, ANT calculates the direction for each evacuee to avoid hazards. In a real evacuation, this direction should be computed and communicated to evacuees via a wired or wireless network. Notably, prior work has often overlooked the possible delays in this communication.
\end{itemize}
Networks and servers used for decision-making may experience congestion, especially in emergencies when decisions and communications are frequent. This congestion can lead to delays in navigation direction updates, packet losses, and errors in decisions due to the use of delayed or obsolete data in decision algorithms \cite{shi2022wireless,dong2022faster}. While previous studies often neglect these effects, this paper specifically evaluates their impact on evacuation time. Furthermore, potential delays and errors stemming from technology, including network packet losses and server failures. 

The simulation framework serves as a comprehensive tool for assessing the intricate dynamics of evacuation scenarios, including the technology-induced delays.

\subsection{System Parameters for the Simulation}

\begin{figure}[h]
	\centering
	\subfigure[]{
		\includegraphics[width=7.8cm,height=4cm]{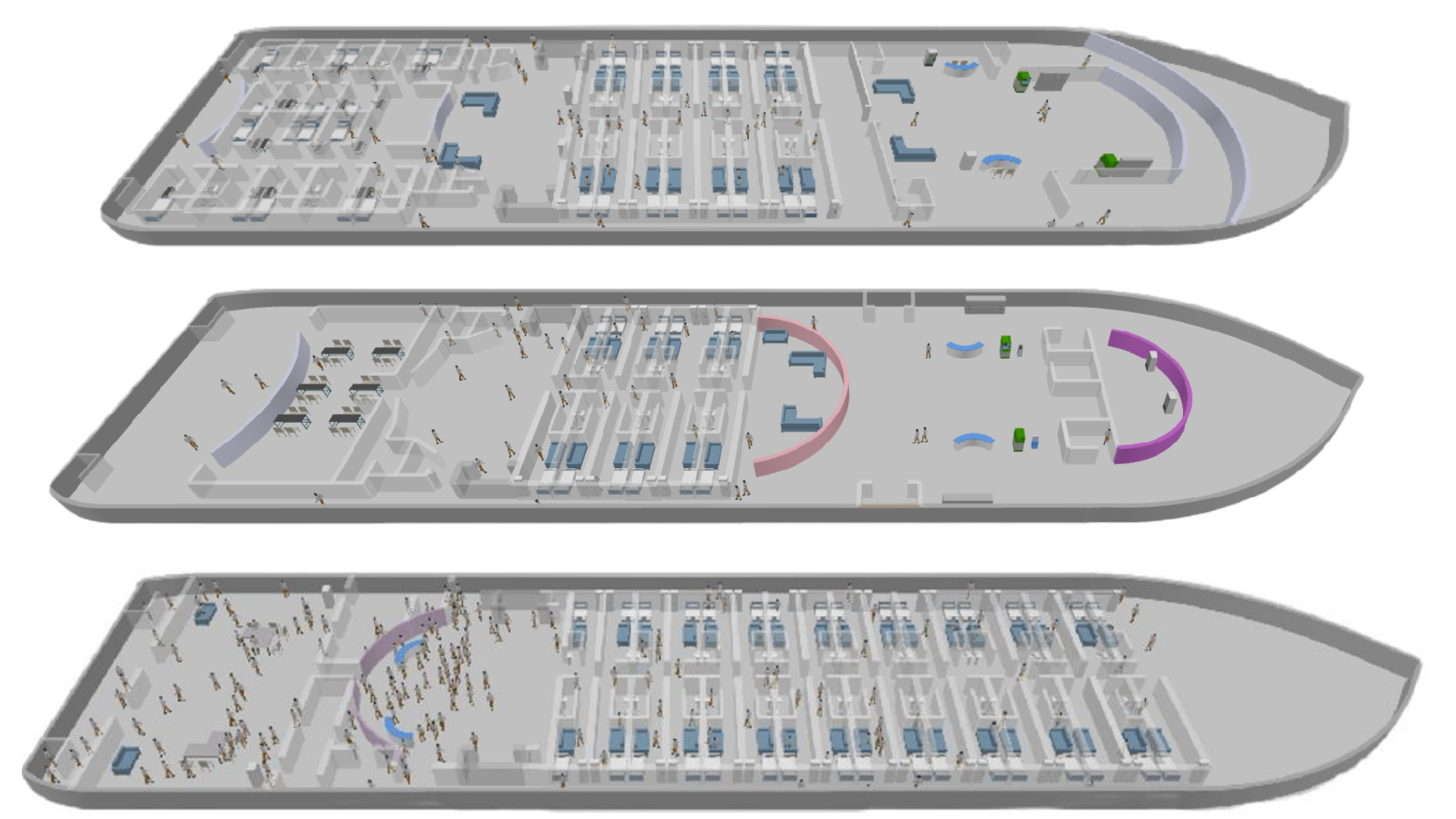}}
	\subfigure[]{
		\includegraphics[width=7.8cm,height=4cm]{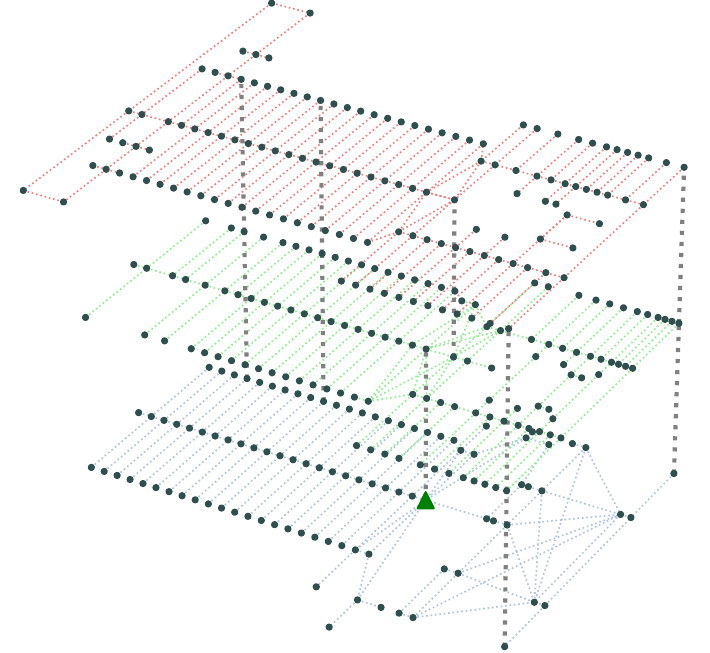}}
	\caption{Schematic description of the layout for the Yangtze Gold 7 Cruise ship over its three passenger floors (second, third, and fourth). This layout is used for simulating the effects that are studied in this paper, i.e., the delays in communicating the guidance information to the evacuees. Here  (a) is the layout of the physical space of the second, third, and fourth floors, and (b) is the evacuation graph model of the physical space.} 
	\label{Fig1} 
\end{figure}

We simulate the indoor environment of the second, third, and fourth floors of the Yangtze Gold 7 Cruise Ship, illustrated in Figure \ref{Fig1}. The simulation includes 346 nodes representing origin points, passageways, connection points between corridors or rooms, and a single exit node for evacuation. The graph also encompasses 600 passageway segments and 5 staircase segments connecting these nodes.

In our simulation setup, the worst-case traversal time across each segment is computed based on the worst-case traversal speed of evacuees, set at $0.067$ meters/second. Conversely, the typical traversal time for each segment reflects the average walking speed on a passenger ship under normal safe conditions, set at $0.67$ meters/second. The total time available for evacuation, denoted as $T_D$, is estimated for evacuees who receive the "evacuate" message:
\begin{equation}
T_D = T_S - T_A - T_{EL},
\end{equation}
where: $T_S$ is the ship's survival duration until capsizing (e.g., 60 minutes),
$T_A$ is the delay between the start of the emergency until the "evacuate" message is received (e.g., 5 minutes),
$T_{EL}$ is the sum of embarkation time and lifeboat launching time for evacuees reaching the exit point.
Guidelines from the Maritime Safety Committee (MSC) \cite{international2016guidelines} specify $T_S = 60$ minutes, $T_A = 5$ minutes, and $T_{EL} = 25$ minutes, resulting in $T_D = 30$ minutes, unless otherwise stated. This simulation framework captures essential parameters for evaluating evacuation strategies in maritime emergency scenarios.

\subsection{The IoT and Information and Communication (IC) Environment}

The IoT and IC environment that is being considered, is composed of:
\begin{enumerate}
	\item A centralized Data Center (DC) that stores the list of passengers and staff who are present on the passenger ship. This list is updated when the passengers or staff board the ship, as well as when anyone leaves the ship for various normal reasons, as well as when a passenger is evacuated through a designated exit.
	\item The DC also computes the {\bf Evacuation Recommendations (ERs)} and movement directions for all the passengers during an evacuation using the ANT algorithm.
	\item A wired communication high-speed Local Area Network (LAN) that runs throughout the ship, with appropriate switching equipment. The LAN is connected to WiFi Hubs and sensors that are placed throughout the ship, including in the cabins, lounges, restaurants, and the passageways and corridors used for normal movement of passengers and staff, and also used for evacuation.
	\item  The cabins and passageways contain Infrared Sensors that can detect human presence in different locations. These infrared sensors do not detect the identity of human beings, but simply their presence. Infrared sensors can also be used to detect high temperatures caused by fire and electrical short circuits.
	\item Passageways and Common Rooms (Restaurants and Lounges) can also contain Video Camera Sensors, that help to count the number of people in a given area, and also help find people who may have lost their way, and are also used for security.
	\item Finally, also connected to the LAN,  are {\em Direction Providers which turn ON during an emergency}. These may be red/white flashing signs that say ``Go Straight", ``Turn Right'', ``Turn Left'', ``Wait Here'', etc. to provide directions
	to evacuees during an emergency. They are activated by the ANT algorithm's ERs 
	that are computed by the DC described in Item 1 above. 
\end{enumerate}
While the individual personal identification of passengers is not considered explicitly for the IoT system
in this paper, systems using wireless ``RFID'' (radio-frequency identity), and  ``wristwatches'' that broadcast the identity of the wearers, can also be used. However, IoT devices with wireless communications may be less reliable inside ships, due to the high metal and steel content of the naval structures, and the resulting absorption, reflection, and refraction of high-frequency radio signals, resulting in high interference and signal attenuation,  which may require sophisticated bandwidth management,  radio-frequency repeaters, and signal processing.

\subsection{Layout of the Simulation Framework}

The schematic diagram of the simulation layout for the three passenger floors of the Yangtze Gold 7 Cruise ship is depicted in Figure \ref{Fig2}. On the right-hand side, the dotted diagram illustrates "dots" representing nodes, which signify locations where passengers may congregate or gather (e.g., cabins, the lobby, and the restaurant). The edges in the diagram depict passageways, stairs, or corridors. 

The simulation incorporates the effect of the inclination angle of the damaged ship, which can change at regular intervals. This inclination can impact the average traversal time across each corridor or staircase, varying with changes in inclination (resulting in shorter or longer traversal times). Thus the ANT provides the best advice that includes the estimate of traversal time, which may change due to the ship's inclination. Each simulation is initiated by placing evacuees (passengers) over the nodes. The simulation is then repeated for 100 rounds, and the initial locations of the evacuees are randomized in each round. This approach ensures robustness and comprehensiveness in evaluating evacuation strategies.

\section{Impact of Delays in Computing and Communications on Passenger Evacuation} \label{SectionIV}

The ANT algorithm provides evacuees a recommendation regarding the next-step hazard-free node that the evacuee should navigate towards the exit, at each node along all evacuation paths on the ship. The recommendation is not based on the evacuee's individual identity, and is generic for all evacuees, and is based on the total minimum delay estimation from the current location to the exit. 

However, emergency conditions are dynamic and subject to rapid changes, and the computer and communication system responsible for computing and forwarding these directions may face congestion during an evacuation. As a result, messages to evacuees may be delayed or lost.

In this section, we analyze the impact of these potential delays caused by performance imperfections and congestion in the Information Technology System (ITS). We introduce the concept of "information lag" ($IL$) for any generic node in the ship evacuation topology. $IL=0$ signifies that each node has received the exact direction recommendation from the ITS system based on the current true location of evacuees. Conversely, $IL=1$ implies that each node provides information to evacuees based on the status of all nodes, derived from the location of evacuees just before their current arrival at the node. Thus, $IL=1$ denotes a delay of one step in the computation and transmission of information.

We also define the "probability of delay" ($PoD$), indicating whether the information lag is $IL=1$ with a probability of $PoD$, or $IL=0$ with a probability of $1-PoD$. $PoD$ is a probabilistic attribute assigned to each node, considering that the delay may vary from node to node due to communication system delays.

In the following analysis, we evaluate the effect of $PoD$ on the evacuation system's performance. Simulation results are obtained by assigning the probability $PoD$ separately for each node in 100 independent simulations under the same initial conditions. The figures presented for each simulation also include the $95\%$ confidence intervals.

\subsection{Evaluation of the Evacuation Times}

\begin{figure}[htbp]
\centering
\subfigure[]{
\includegraphics[width=8cm,height=5cm]{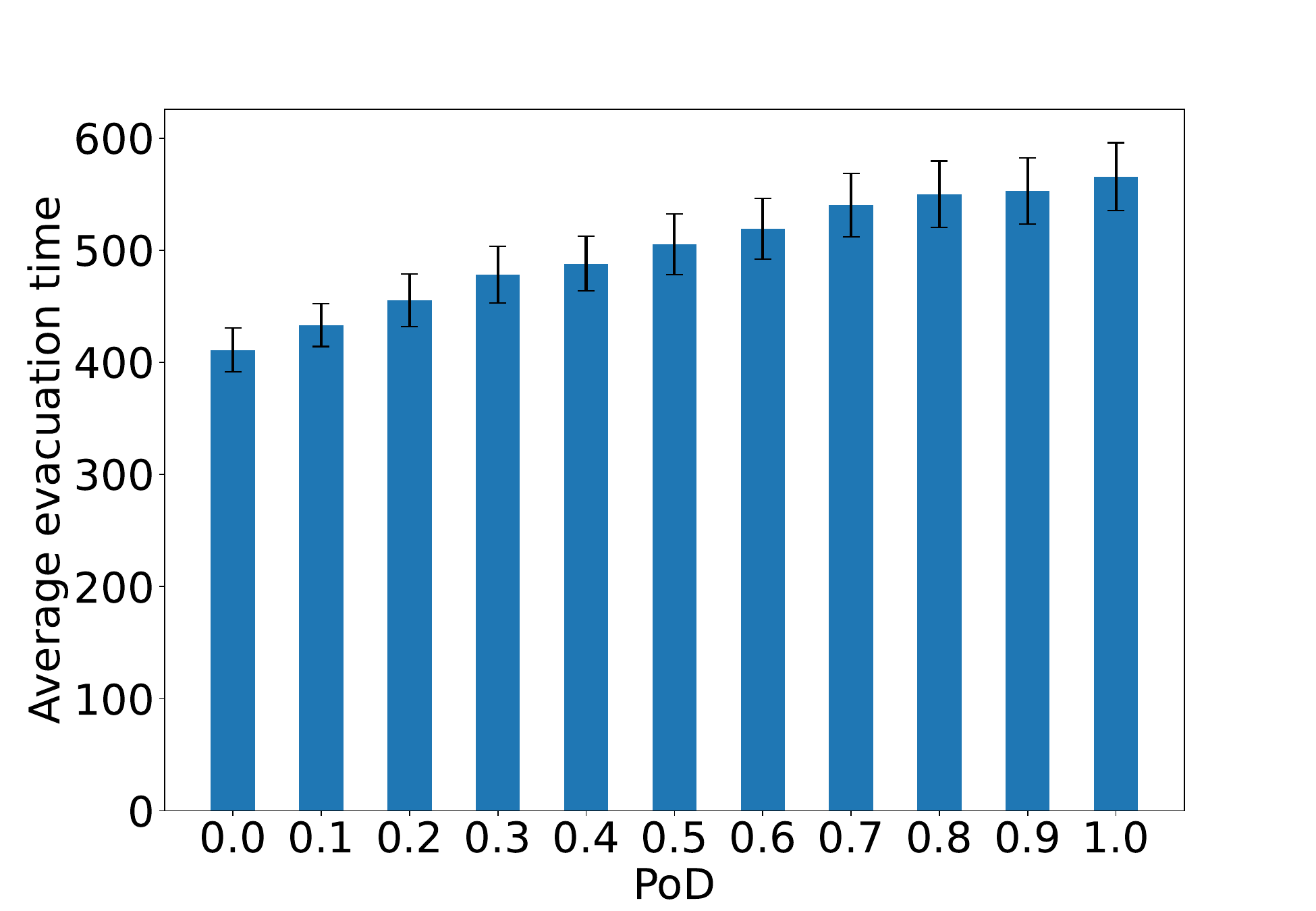}
}
\subfigure[]{
\includegraphics[width=8cm,height=5cm]{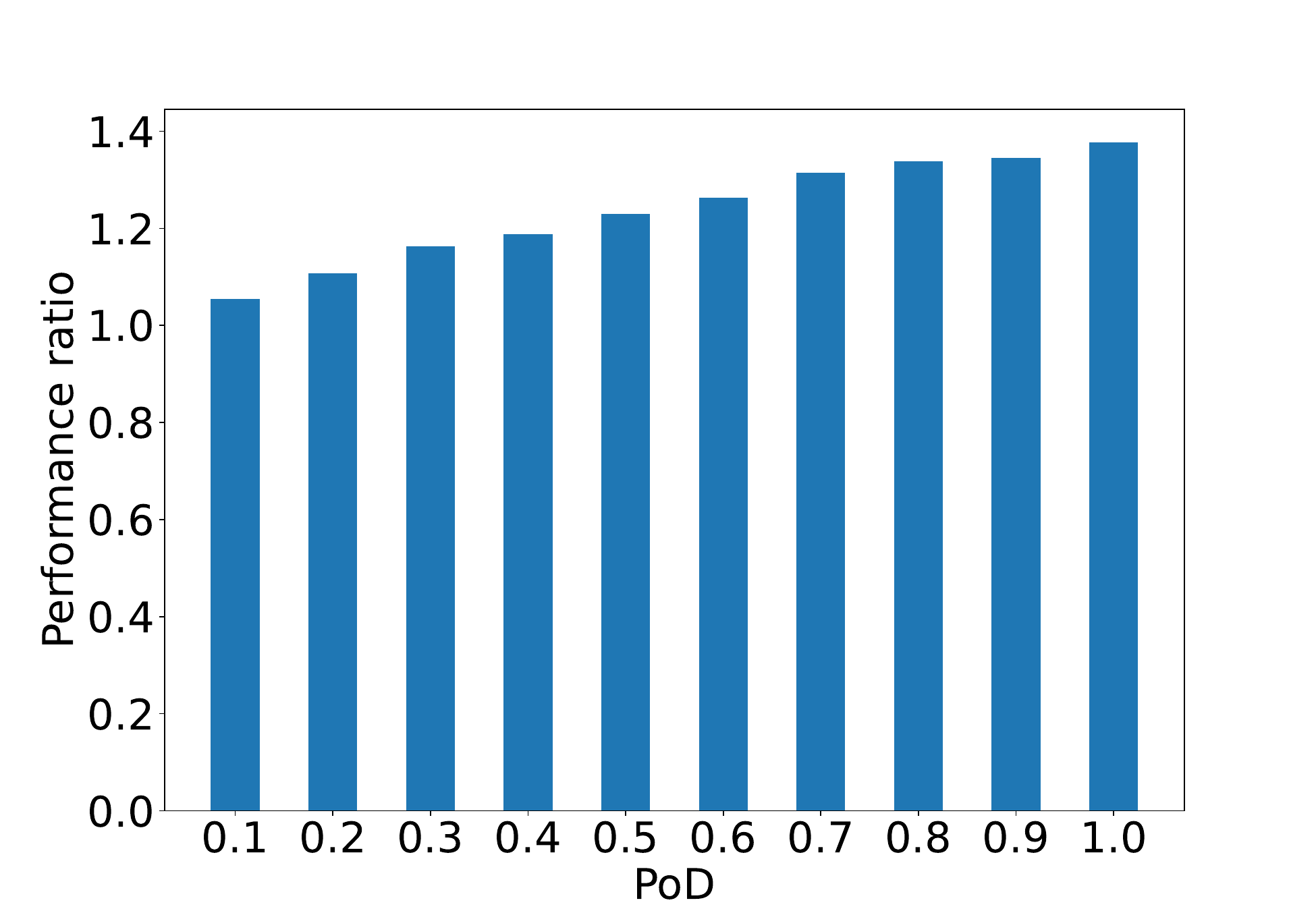}
}
\caption{The average evacuation time, in seconds, from all $346$ nodes to the exit (above), as a function of $PoD$ (x-axis), and (below) the performance ratio in average evacuation time as compared to the ideal case of $PoD=0$. Averages are over all nodes for $100$ distinct independent simulations, with the standard deviation (the black bars) for the evacuation time.}
\label{Fig2}
\end{figure}

We first determine the average evacuation time from all 346 nodes concerning the ideal case with $IL=0$. Figure \ref{Fig2} (a) illustrates the average evacuation time from all 346 starting nodes to the exit as a function of $PoD$. Figure \ref{Fig2} (b) presents the performance ratio of average evacuation time for various $PoD$ values relative to the average for the ideal case of $PoD=0$. The averages are taken over all nodes and based on 100 distinct independent simulations. The standard deviation is also shown as black bars for the evacuation time from all 346 nodes. These curves demonstrate a quasi-linear increase in average evacuation delay as $PoD$ increases to 0.5, and a more gradual increase for higher values. When $PoD=1$ for all nodes, the average evacuation time over all nodes is $50\%$ higher than $PoD=0$, where the ITS provides up-to-date information to all nodes.

\begin{figure}[htbp]
\centering
\subfigure[]{
\includegraphics[width=8.1cm,height=5cm]{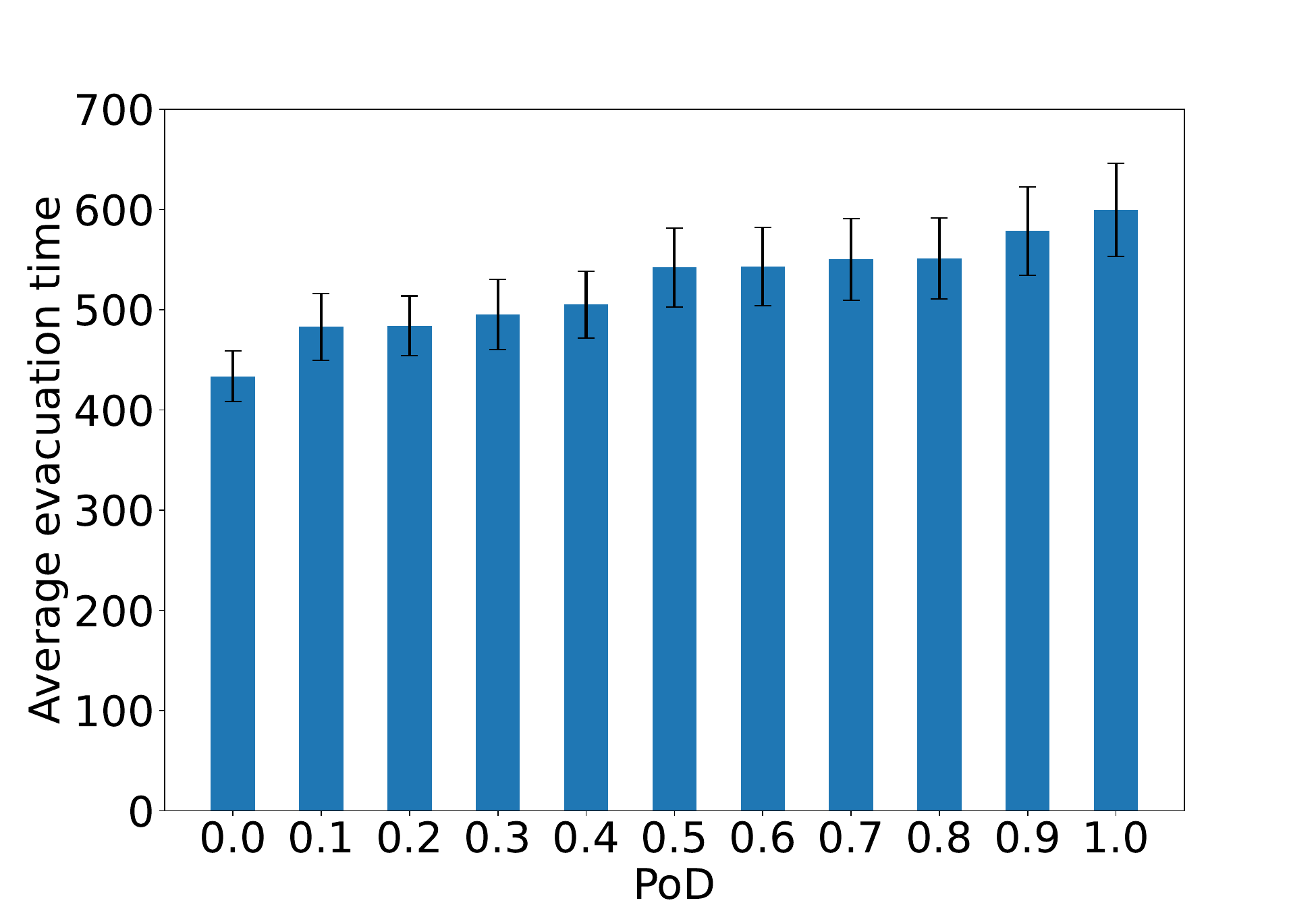}
}
\subfigure[]{
\includegraphics[width=8.1cm,height=5cm]{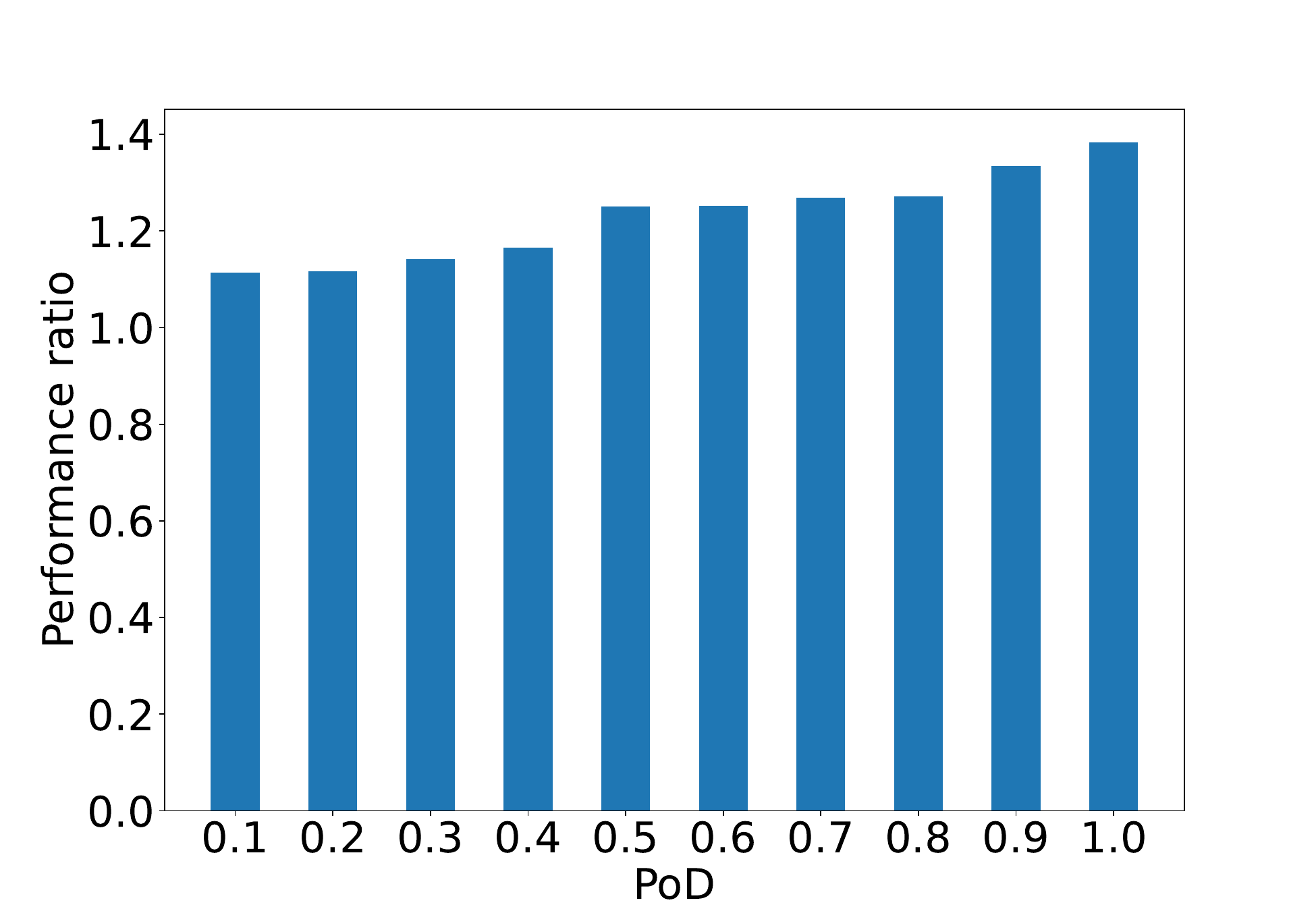}
}
\caption{The average evacuation time in seconds, taken by passengers in cabins to the exit (left), as a function of $PoD$ (x-axis), and (right) the performance ratio in average evacuation time as compared to the ideal case of $PoD=0$. Averages are taken over all passengers starting from cabins for $100$ distinct independent simulations, with the standard deviation (the black bars) for the evacuation time.}
\label{Fig3}
\end{figure}

Additionally, we perform a set of simulations to evaluate the average evacuation time for passengers located in cabins. Figure \ref{Fig3} (above) illustrates the average evacuation time in seconds, taken by passengers originating from cabins, affected by different probabilities $PoD$. The performance ratio in average evacuation time for passengers in cabins, relative to the case $PoD=0$, is shown in Figure \ref{Fig3} (b). It is evident that the evacuation of passengers from cabins worsens with the increase in $PoD$. 

\begin{figure}[htbp]
\centering
\subfigure[]{
\includegraphics[width=8.1cm,height=5cm]{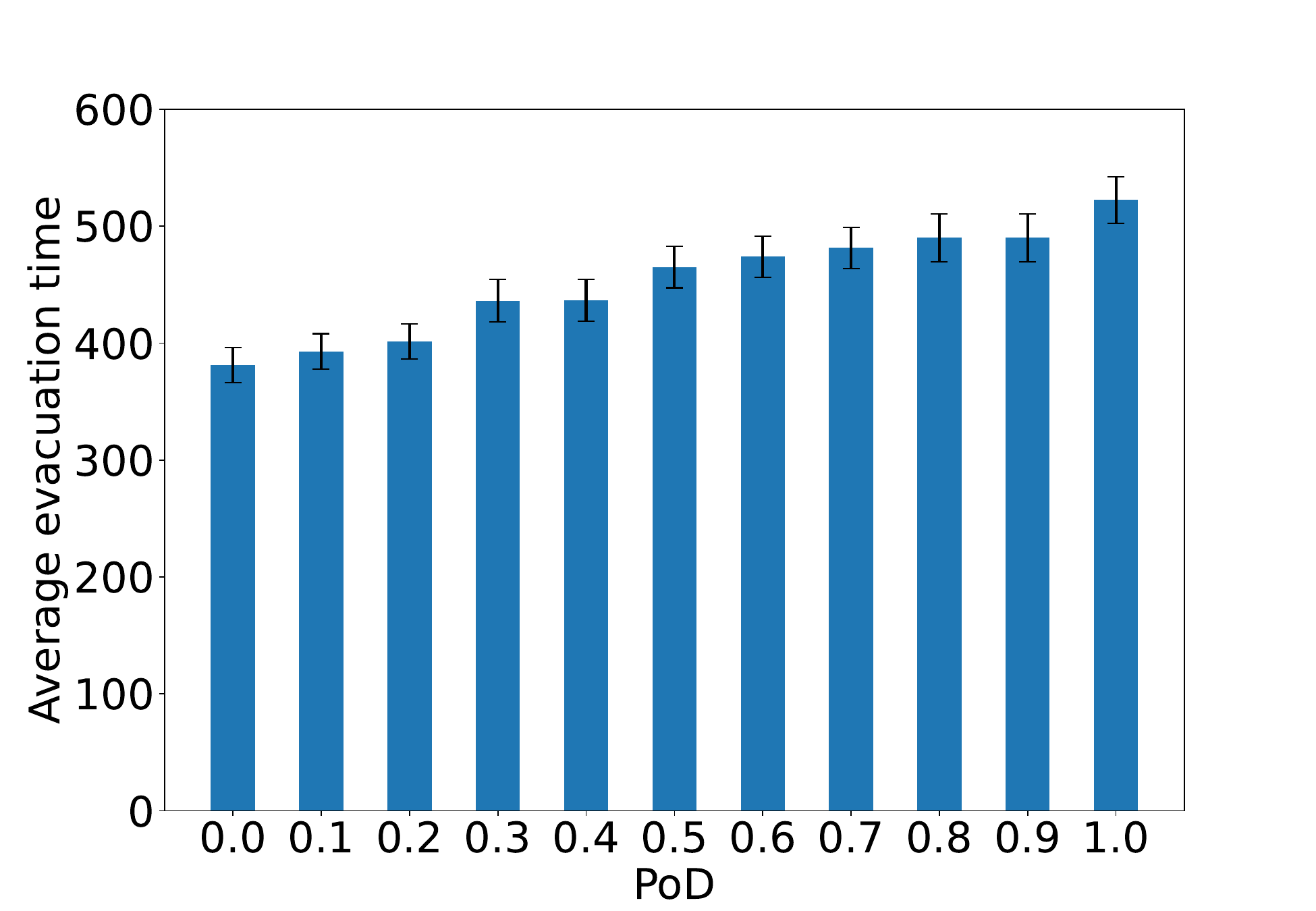}
}
\subfigure[]{
\includegraphics[width=8.1cm,height=5cm]{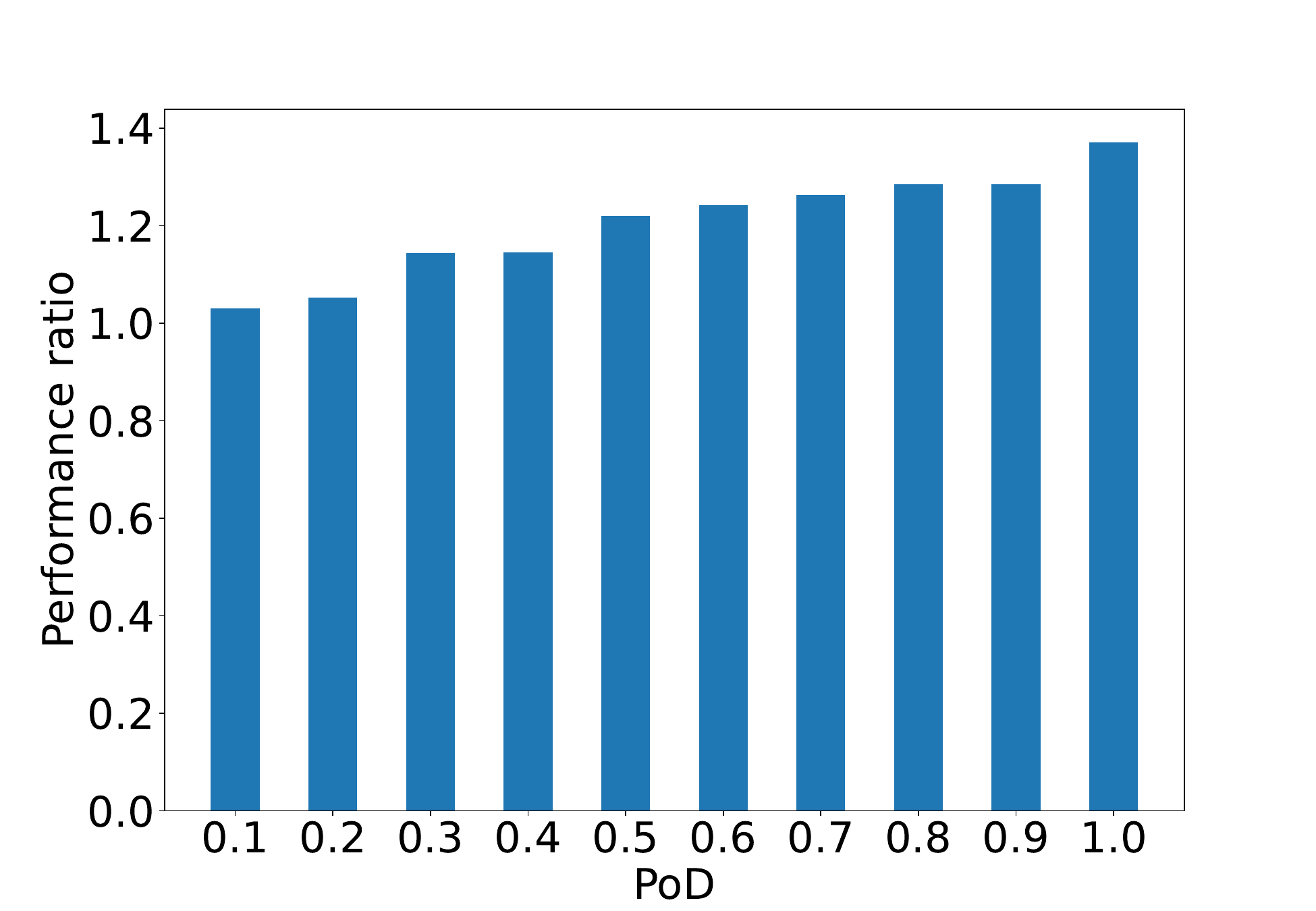}
}
\caption{The average evacuation time in seconds, taken of passengers in the restaurant to the exit (left), as a function of $PoD$ (x-axis), and (right) the performance ratio in average evacuation time as compared to the ideal case of $PoD=0$. Averages are taken over all passengers in the restaurant for $100$ distinct independent simulations, with the standard deviation (the black bars) for the evacuation time.}
\label{Fig4} 
\end{figure}

Moreover, we investigate the average evacuation time for passengers located in the restaurant. Figure \ref{Fig4} (a) illustrates the average evacuation time of passengers within the restaurant. Notably, we observe that IL has a milder impact on the evacuation time of restaurant patrons compared to passengers in cabins, particularly when $PoD$ is low, such as $PoD=0.1$ and $PoD=0.2$ 

\begin{figure}[htbp]
	\centering
       \subfigure[]{
       \includegraphics[width=8.57cm,height=5cm]{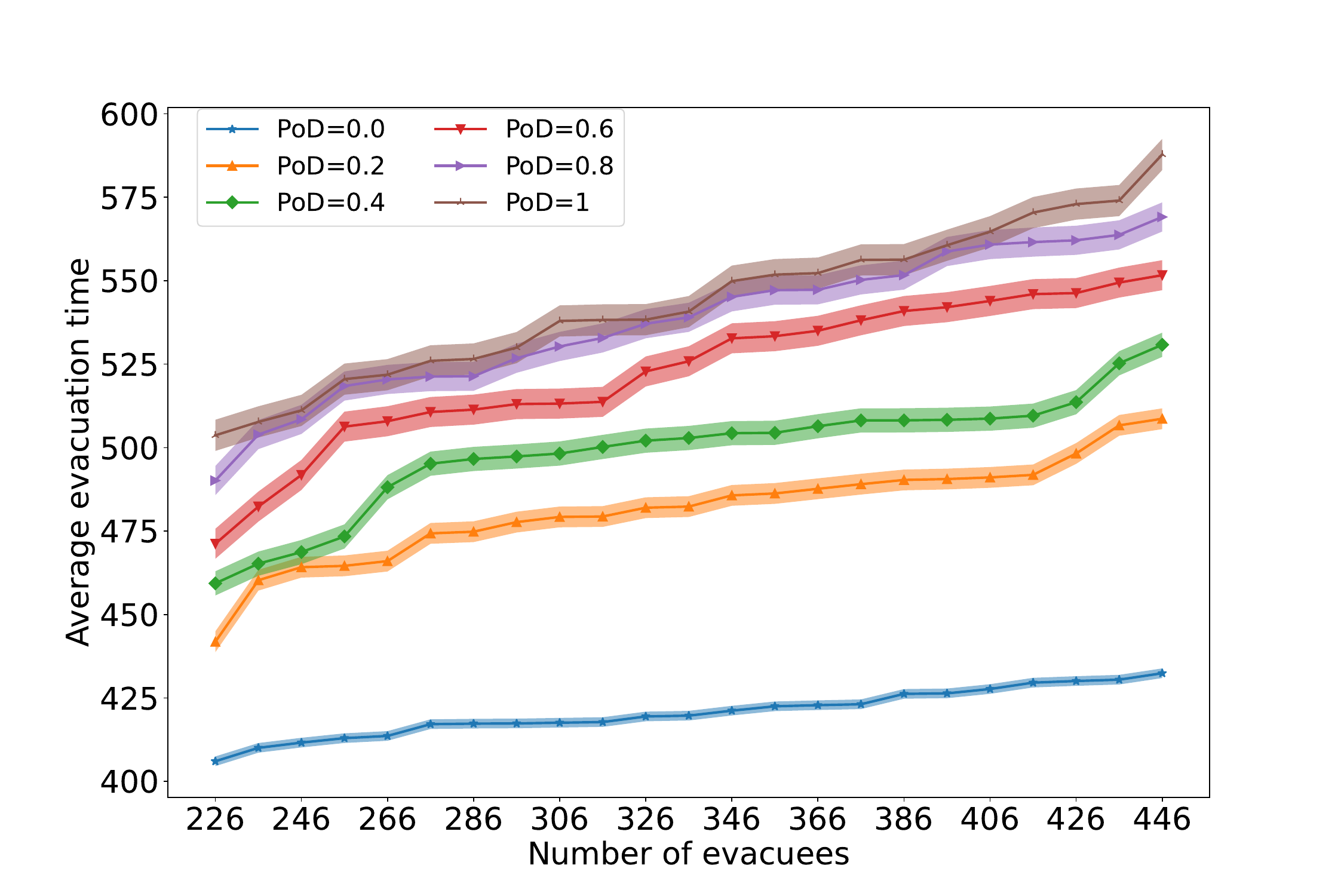}
                   }
      \subfigure[]{
      \includegraphics[width=8.57cm,height=5cm]{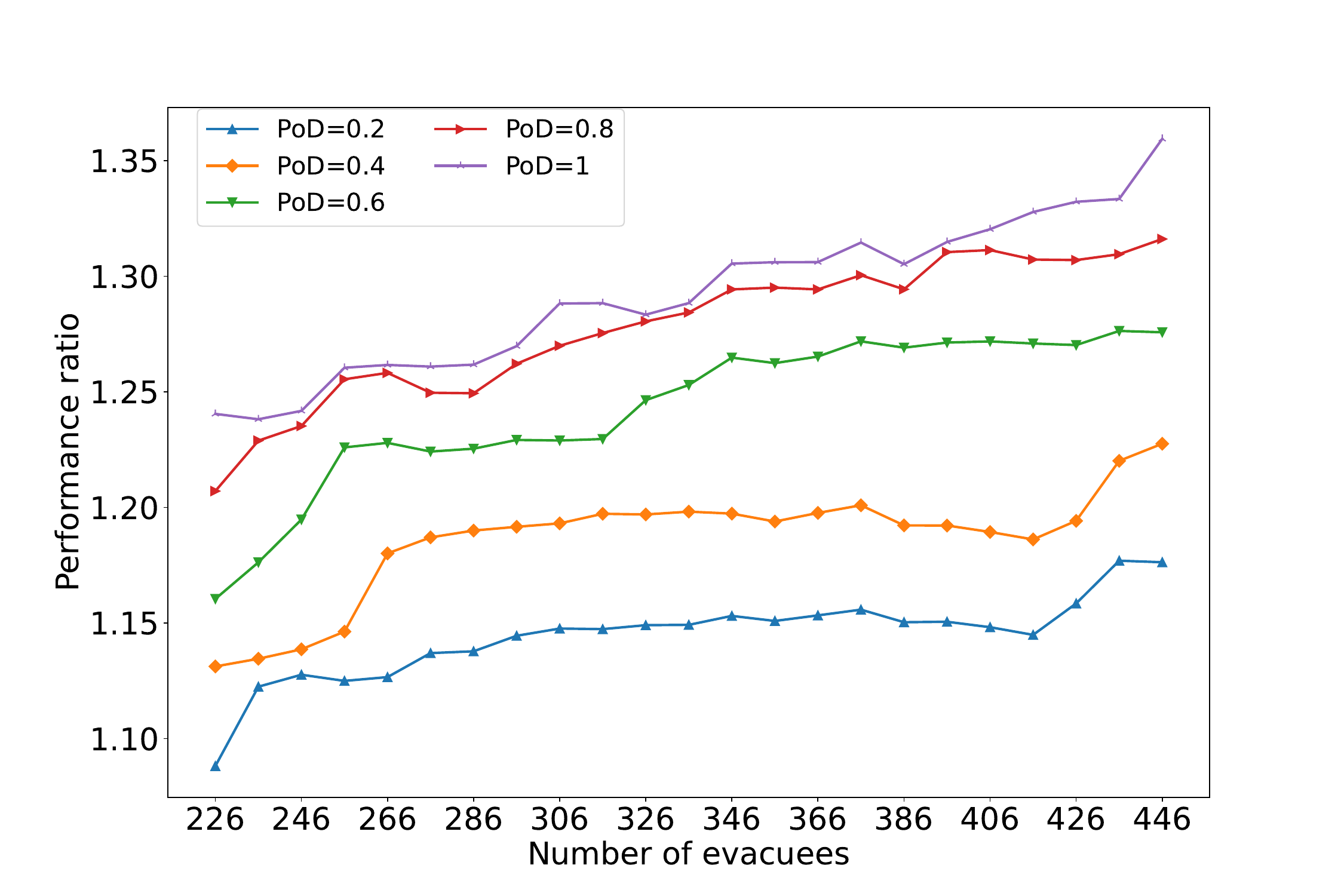}
                   }  
     \caption{The average evacuation time in seconds, with a $95\%$ confidence interval, taken by different numbers of passengers where half of them originate  in the cabins,  while the other half start from  the restaurant (above), and (below) the performance ratio in average evacuation time as compared to the ideal case w$PoD=0$.} 
    \label{Fig5} 
\end{figure}

Additionally, other simulations assess the average evacuation time for varying numbers of passengers, influenced by different $PoD$ values. In this simulation set, half of the passengers are randomly distributed in cabins, while the remaining half occupy the restaurant. Notably, the cabin capacity is limited to two passengers at most. As depicted in Figure \ref{Fig5} (a), the average evacuation time demonstrates an increase with a growing number of passengers. This escalation is primarily attributed to heightened waiting times resulting from more pronounced congestion caused by the evacuation of a larger number of passengers. Moreover, we observe that the average evacuation time rises with $PoD$, irrespective of the number of passengers. Figure \ref{Fig5} (b) presents the performance ratio in average evacuation time for varying passenger numbers under different probabilities of $PoD$. The performance ratio consistently exceeds $1$, indicating that a larger $PoD$ leads to an extended evacuation duration, with a more pronounced impact observed for higher $PoD$ values.

\section{Conclusions and Further Research} \label{SectionVII}
Ensuring reliable emergency evacuation is vital, and IoT systems to support evacuation and guide evacuees 
have been the object of much research  
\cite{wang2016send, Hao}. However, challenges arise during emergencies due to congestion and damage to the IoT system supporting an evacuation, as well as to damages in the built environment.

This paper has focused on the evacuation of passengers from a cruise ship during an emergency, and has considered the ICT and IoT system that is needed to enable the timely evacuation of the passengers crew and staff. In particular, we have focused on the impact of the IoT through sensor networks \cite{pan2006emergency,li2009ern}, and the computing and communication system, on the effectiveness of the navigation advice provided to passengers. This study represents the first examination of the significant side effects of ICT and IoT performance influencing emergency evacuation in practical ship scenarios.  With extensive simulations, the paper investigates the impact of the delay in receiving correct movement instructions by the passengers, on the evacuation performance during a ship emergency. The simulations incorporate parameters from the real passenger cruise ship "Yangtze Gold 7" and its evacuation system, and assess the repercussions of information delays reaching human evacuees during evacuation.

In future research, it will be useful to proactively incorporate the delaying factors, and design innovative decentralized emergency navigation systems that may locally pre-compute and store in a distributed manner the needed advisory data at key system intermediate nodes, and combine centralized decisions with individual evacuee decision aids.
\bibliographystyle{IEEEtran}
\bibliography{refs} 

\end{document}